\documentclass[prl,aps,twocolumn,showpacs]{revtex4}

\usepackage{color,amsmath,graphicx,epsfig}

\begin{document}

\def\pd#1#2{\frac{\partial #1}{\partial #2}}
\def\bk{{\bf k}}
\def\bx{{\bf x}}

\title{\bf Condensation of Classical Nonlinear Waves in a Two-Component System}
\author {Hayder Salman and Natalia G.\ Berloff}
\affiliation {Department of Applied Mathematics and Theoretical Physics,
University of Cambridge,  Cambridge, CB3 0WA, UK\\
}

\begin {abstract}
We study the formation of large-scale coherent structures (a condensate) 
for a system of two weakly interacting classical waves. Using the coupled 
defocusing nonlinear Schr\"{o}dinger (NLS) equations as a representative model,
we focus on condensation in the phase mixing regime. We employ weak
turbulence theory to provide a complete thermodynamic description of
the classical condensation process. We show that the temperature and
the condensate mass fractions are fully determined by the total
number of particles in each component and the initial total energy.
Moreover, we find that, at higher energies, condensation can occur
in only one component.
The theory presented provides excellent agreement with
results of numerical simulations obtained by directly integrating the dynamical
model.

\end{abstract}
\maketitle

Many classical systems in nature reveal the emergence of large scale
coherent structures from a background irregular field characterised
by small-scale fluctuations. Examples of systems that exhibit such
behaviour include classical turbulence, nonlinear optics,
superfluids, ultracold gases and Bose-Einstein Condensates (BECs),
and the formation of the early universe. In certain regions of the
parameter space, a large sub-class of these systems can be described
by a system of weakly nonlinear dispersive waves. A universal
equation that governs the evolving field in such scenarios is then
given by the Nonlinear Schr\"{o}dinger (NLS) equation. The process
of self-organisation in the focusing NLS equation has been studied
in \cite{ZakDya}. It was found that a large-scale solitary wave
tends to emerge from a sea of small-scale turbulent fluctuations. In
this work, we concentrate on the defocusing NLS equation. This
equation has been receiving increasing attention due to the
experimental advances in BECs. In this context, the defocusing NLS
equation corresponds to the Gross-Pitaevskii (GP) equation \cite{GP}
of a homogeneous Bose gas. The GP equation has long been used as a
model of a weakly interacting Bose gas at zero temperature. More
recently, it has been argued \cite{svist} that the GP equation can
be used to model the long wavelength part of the spectrum of a BEC
at finite temperatures. Numerical simulations conducted within this
framework \cite{burnett}, \cite{bs} have indeed confirmed this,
revealing the ability of the model to capture the formation of a
condensate from an initially turbulent state.

With the rapid developments being made in experimental techniques, it is now
possible to realize multi-component BECs formed by the simultaneous trapping
and cooling of atoms in distinct spin or hyperfine levels \cite{mixture1} or
different atomic species \cite{mixture2}. The finite temperature dynamics of
such Bose gas mictures is then goverened by a system of coupled NLS equations.
While such systems are of of interest in
their own right, they also
serve as idealised models to study symmetry-breaking phase transitions that are
believed to have occurred in the early evolution of the universe. A specific
example is given by the Kibble-Zurek mechanism \cite{kz} of the formation of
topological defects following the rapid quench of the system below the point of
second-order phase transitions. This scenario would correspond to the formation of
cosmological {\em vortons} and {\em springs} that are analogous to the
vortex ring-slaved wave and vortex ring-vortex ring complexes of BECs
\cite{berloff05b}. In addition to these physical examples, the coupled 
NLS equations are also encountered in the study of optical fibres and 
electromagnetic waves \cite{Manakov74}. Given the universality of these equations in the nonlinear
sciences, an accurate thermodynamic description of the condensation process in such a
system could have significant implications in various branches of physics.

In this Letter, we will generalise the results presented in
\cite{josserand-05}, that describe the condensation process of a
one-component system, to a two-component system. We note that
two-component systems tend to show a broad class of qualitatively
different behavior depending on the relative strengths of the
intercomponent and intracomponent interactions. This can lead
to contrasting regimes of condensation: the phase mixing regime and
the phase separation regime \cite{hualin}. Consistent with our
original assumptions of weakly interacting systems, we will focus
exclusively on the phase mixing regime.

We begin by considering the scenario of a system of two weakly interacting 
classical waves (e.g.\ Bose gases consisting of the
same atoms in different spins) that have been rapidly cooled below the 
transition temperature. Their evolution from the
resulting strongly nonequilibrium initial state is then 
described by the coupled NLS equations given by
\begin{eqnarray}
i \partial_t \psi_1 &=& -\nabla^2 \psi_1 + |\psi_1|^2 \psi_1 + \alpha|\psi_2|^2 \psi_1,\nonumber\\
i \partial_t \psi_2 &=& -\nabla^2 \psi_2 + |\psi_2|^2 \psi_2 + \alpha|\psi_1|^2 \psi_2 ,
\label{eqn_psi}
\end{eqnarray}
where $\psi_1$ and $\psi_2$ are complex-valued classical fields corresponding 
to each component, and $\alpha$ is the intracomponent coupling constant. 
For the phase mixing regime, 
we require $0 < \alpha < 1$. The dynamics governed by the above equations will
conserve the total mass (number of particles) given by
$N_1 = \int |\psi_1|^2 d\mathbf{x}$ and $N_2 = \int |\psi_2|^2
d\mathbf{x}$. In addition, the total energy (Hamiltonian) of the
coupled system
\begin{equation}
H=\int \biggl[ \sum_{i=1}^2\{|\nabla \psi_i|^2 + \frac{1}{2}
|\psi_i|^4\}+ \alpha |\psi_1|^2|\psi_2|^2 \biggr]\, dV.
\label{eqn_energy}
\end{equation}
will be conserved. Without loss of generality we shall assume that $N_2 \le N_1$.

Despite the formal reversibility of the above Hamiltonian system,
the evolution of the nonlinear waves $\psi_1$ and $\psi_2$ is
nonintegrable giving rise to an effective diffusion in phase space.
This results in an {\em irreversible} evolution to thermal
equilibrium. By invoking the random phase approximation (assumption
of quasi-Gaussian statistics), it is possible to derive closed
irreversible kinetic equations that describe the evolution of the
system using Weak Turbulence Theory (WTT) \cite{ZakhBook}. For a
homogeneous system, we accomplish this by expressing the order
parameters in terms of their Fourier transforms $\psi_1 =
\frac{1}{(2\pi)^{3/2}} \int a_{\mathbf{k}}(t) e^{i\mathbf{k} \cdot
\mathbf{x}} d\mathbf{k}$, $\psi_2 = \frac{1}{(2\pi)^{3/2}} \int
b_{\mathbf{k}}(t) e^{i\mathbf{k} \cdot \mathbf{x}} d\mathbf{k}$.
Substituting into Eq.\ (\ref{eqn_energy}),
we can derive expressions for the spectral number densities $\left<
a_{\mathbf{k}_1} a_{\mathbf{k}_2}^* \right> = n_{1}
\delta(\mathbf{k}_1 - \mathbf{k}_2)$; $\left< b_{\mathbf{k}_1}
b_{\mathbf{k}_2}^* \right> = l_{1} \delta(\mathbf{k}_1 -
\mathbf{k}_2)$. Provided the nonlinearity in the system is
sufficiently weak (i.e.\ $N_1/V \ll 1$; $N_2/V \ll 1$; $\alpha \ll
1$, where $V$ is the volume of the system), we can derive the
kinetic equation
\begin{eqnarray}
&&\partial_t {n}_{k} = \frac{4\pi}{(2\pi)^6}\int \biggl(
\biggl[(n_k+n_1)n_2n_3 - n_k n_1 (n_2+n_3)\biggr] \nonumber \\
&& + \alpha^2 \biggl[(n_k+l_1)l_2n_3 - n_k l_1
(l_2+n_3)\biggr]\biggr) \label{eqn_kn1} \\
&& \times \delta(\mathbf{k} + \mathbf{k}_1 - \mathbf{k}_2 -
\mathbf{k}_3) \delta(k^2+k_1^2-k_2^2-k_3^2) d\bk_{1} d\bk_{2} d\bk_{3}. \nonumber
\end{eqnarray}
Similarly, an equation for $l_k$ can be derived and follows directly
from above by the symmetry in Eq.\ (\ref{eqn_psi}). These equations
conserve $N_1 = V \int n_{\mathbf{k}}(t) d\mathbf{k}$, $N_2 = V \int
l_{\mathbf{k}}(t) d\mathbf{k}$, and the kinetic energies $E_1 = V
\int k^2 n_{\mathbf{k}}(t) d\mathbf{k}$, $E_2 = V \int k^2
l_{\mathbf{k}}(t) d\mathbf{k}$ of each component. They admit two formal
equilibrium solutions; the first corresponding to a uniform
distribution $n_{\mathbf{k}}^{\mathrm{eq}} = c_1$,
$l_{\mathbf{k}}^{\mathrm{eq}} = c_2$, and the second given by the
Rayleigh-Jeans (RJ) distribution
\begin{eqnarray}
n_{\mathbf{k}}^{\mathrm{eq}} = \frac{T}{k^2-\mu_1}, \;\;\;\;\;
l_{\mathbf{k}}^{\mathrm{eq}} = \frac{T}{k^2-\mu_2}.
\label{eqn_RJ}
\end{eqnarray}
Here, $T$ is the thermodynamic temperature, and $\mu_1$ and $\mu_2$
are the chemical potentials. Equation (\ref{eqn_kn1})
satisfies a H-theorem for entropy growth which implies that the RJ
distribution will be realized in practice. However, Eq.\
(\ref{eqn_RJ}) provides only a formal solution since it leads to
non-convergent expressions for $N_1$, $N_2$, and the kinetic
energies $E_1$, $E_2$ as $k \rightarrow \infty$. We recall that, for BECs,
Eq.\ (\ref{eqn_psi}) is valid in the limit of large occupation numbers
where a semi-classical description is valid. When $n_{\mathbf{k}}
\sim 1$ and $l_{\mathbf{k}} \sim 1$, Eq.\ (\ref{eqn_psi}) begins to
break down and a full quantum mechanical treatment of the problem
becomes necessary.  To regularise the ultra-violet catastrophe, we
introduce a cut-off $k_c$ such that $n^{\mathrm{eq}}(|k_c|)
> 1$, $l^{\mathrm{eq}}(|k_c|) > 1$. This cut-off does not affect the
equilibrium state provided a sufficiently large number of modes can
be represented classically \cite{burnett,davis}. The reason is that
a full quantum mechanical description corresponds to a
grand-canonical ensemble with fluctuations in particle number and
energy. However, for sufficiently many modes, such fluctuations will
be small and we can introduce the above truncation to reduce the
system to a microcanonical ensemble where the number of particles
and the energy are conserved.

The RJ distributions corresponding to Eq.\ (\ref{eqn_kn1}) are only
valid at sufficiently high energies when no condensate is present.
At sufficiently low energies, Eq.\ (\ref{eqn_kn1}) breaks down very
rapidly giving way to the formation of a condensate as elucidated in
numerical simulations for a one-component system
\cite{burnett,bs,josserand-05} and a two-component system \cite{bc}.
In the simplest scenario, condensates with zero wavenumbers are
formed and are associated with the uniform states provided we are in
the phase mixing regime. If the condensates that form are strong in
the sense that $(N_1-n_o)/N_1 \ll 1$, $(N_2-l_o)/N_2 \ll 1$ ($n_o =
|a_o|^2$; $l_o = |b_o|^2$ are the occupation numbers of the
condensates in components 1 and 2, respectively), one can describe
the nonlinear dynamics at these later times by considering the
evolution of small quasiparticle perturbations around the
condensates. For our two-component system, we accomplish this by
introducing the ansatz $a_k = [\sqrt{n_o} \delta(k) +
\widetilde{a}_k(t)]e^{-i n_o t}$, $b_k = [\sqrt{l_o} \delta(k) +
\widetilde{b}_k(t)]e^{-i l_o t}$. Upon substituting these
expressions into the Fourier-transform representation of Eq.\
(\ref{eqn_energy}), we introduce a transformation to diagonalize
terms in the Hamiltonian that are quadratic in $\widetilde{a}_k$ and
$\widetilde{b}_k$. A generalisation of Bogoliubov's transformation
\cite{bog} to diagonalise the Hamiltonian in two-component systems
was given in \cite{timmermans}. 
To this end, we
introduce the canonical variables $\mathcal{N}=(\xi_{\mathbf{k}},
\eta_{\mathbf{k}})^T$ which are related to the original variables
$\mathcal{A}=(\widetilde{a}_{\mathbf{k}},\widetilde{b}_{\mathbf{k}})^T$
through the relation
\begin{eqnarray}
\left( \begin{array}{c}
\mathcal{A} \\
\mathcal{A}^{\dag}
\end{array} \right) =
\left( \begin{array}{cc}
\mathcal{U^+} & \mathcal{U^-} \\
\mathcal{U^-} & \mathcal{U^+}
\end{array} \right)
\left( \begin{array}{c}
\mathcal{N} \\
\mathcal{N}^{\dag}
\end{array} \right),
\end{eqnarray}
where $\mathcal{N}^{\dag}=(\xi_{\mathbf{-k}}^*,
\eta_{\mathbf{-k}}^*)^T$, and
$\mathcal{A}=(\widetilde{a}_{\mathbf{-k}}^*,\widetilde{b}_{\mathbf{-k}}^*)^T$.
To preserve the properties of the Poisson bracket in the new basis,
the transformation must satisfy the condition
$\mathcal{U^+}\mathcal{U^+}^T-\mathcal{U^-}\mathcal{U^-}^T=\bf{I}$. A
transformation that satisfies this condition and diagonalises the
quadratic term is obtained when the elements $u_{ij}^{\pm}$ of the
$2 \times 2$ transformation matrices $\mathcal{U}^{\pm}$ are given
by
\begin{eqnarray}
\mathcal{U}^{\pm} = \left( \begin{array}{cc}
\frac{\Gamma_{k}^{+^2} \pm 1}{2\Gamma_{k}^+} \cos \gamma_k &
-\frac{\Gamma_{k}^{-^2} \pm 1}{2\Gamma_{k}^-} \sin \gamma_k \\
\frac{\Gamma_{k}^{+^2} \pm 1}{2\Gamma_{k}^+} \sin \gamma_k &
\frac{\Gamma_{k}^{-^2} \pm 1}{2\Gamma_{k}^-} \cos \gamma_k
\end{array}
\right).
\end{eqnarray}
 $\Gamma_{k}^{\pm}=\sqrt{k^2/\Omega^{\pm}}$ denotes the ratios of
the dispersion relations, where $\Omega^{\pm} = \sqrt{k^4+c^{\pm}k^2}$
and $c^{{\pm}^2} = [n_o + l_o \pm \sqrt{(n_o-l_o)^2+4\alpha^2 n_o l_o}]/V; \;
\sin \gamma_k = \sqrt{\frac{1}{2}\left[1-
\frac{1-r}{\sqrt{(1-r)^2+4z^2}} \right]}; \cos \gamma_k = \sqrt{\frac{1}{2}
\left[1+\frac{1-r}{\sqrt{(1-r)^2+4z^2}} \right]}$, where $r = l_o/n_o$
and $z = \alpha \sqrt{l_o/n_o}$. The resulting expression for the Hamiltonian 
leads to kinetic equations for the canonical (quasiparticle) densities
$\left< \xi_{\mathbf{k}_1} \xi_{\mathbf{k}_2}^* \right> =
\widetilde{n}_{1} \delta(\mathbf{k}_1 - \mathbf{k}_2)$; $\left<
\eta_{\mathbf{k}_1} \eta_{\mathbf{k}_2}^* \right> = \widetilde{l}_{1}
\delta(\mathbf{k}_1 - \mathbf{k}_2)$ which are
given by
\begin{eqnarray}
\partial_t \widetilde{n}_{k} &=& \pi \int \biggl(
\left[ R_{k12} - R_{1k2} - R_{2k1} \right] + \left[ S_{1k2} +
S_{12k} \right] \nonumber \\
&& \;\;\;\;\;\; + \left[ T_{k12} - T_{1k2} \right] + U_{k12} +
V_{21k}
\biggr) d\bk_{1} d\bk_{2}, \label{kn2} \\
R_{k12} &=& \Delta_{k12}^{(1)} \left[ \widetilde{n}_1
\widetilde{n}_2 - \widetilde{n}_k \widetilde{n}_1 - \widetilde{n}_k
\widetilde{n}_2 \right]
\delta(\Omega_{k}^+ - \Omega_{1}^+ - \Omega_{2}^+), \nonumber \\
S_{1k2} &=& \Delta_{1k2}^{(2)} \left[ \widetilde{n}_k \widetilde{n}_2
- \widetilde{l}_1 \widetilde{n}_k - \widetilde{l}_1 \widetilde{n}_2
\right]
\delta(\Omega_{1}^- - \Omega_{k}^+ - \Omega_{2}^+), \nonumber \\
T_{k12} &=& \Delta_{k12}^{(3)} \left[ \widetilde{n}_1 \widetilde{l}_2 -
\widetilde{n}_k \widetilde{n}_1 - \widetilde{n}_k \widetilde{l}_2
\right]
\delta(\Omega_{k}^+ - \Omega_{1}^+ - \Omega_{2}^-), \nonumber \\
U_{k12} &=& \Delta_{k12}^{(4)} \left[ \widetilde{l}_1 \widetilde{l}_2 -
\widetilde{n}_k \widetilde{l}_1 - \widetilde{n}_k \widetilde{l}_2
\right]
\delta(\Omega_{k}^+ - \Omega_{1}^- - \Omega_{2}^-), \nonumber \\
V_{21k} &=& \Delta_{21k}^{(5)} \left[ \widetilde{l}_1 \widetilde{n}_k
- \widetilde{l}_2 \widetilde{l}_1 - \widetilde{l}_2 \widetilde{n}_k
\right] \delta(\Omega_{2}^- - \Omega_{1}^- - \Omega_{k}^+),
\nonumber
\end{eqnarray}
where $\Delta_{k12}^{(i)} =
C_{k12}^{(i)}\delta(\mathbf{k}-\mathbf{k}_1-\mathbf{k}_2)$ and
$C^{(i)}$ denote coefficients that will, in general, depend on
$u_{ij}^{\pm}$, $n_o$, $l_o$ and $\alpha$. Since we are only
interested in equilibrium solutions, their precise form is not too
important. The equation for $\widetilde{l}_{\mathbf{k}}$ follows by
symmetry of the dynamical equation. These kinetic equations are now
given by resonant three-wave interactions and have a one parameter
family of solutions given by
$\widetilde{n}_{\mathbf{k}}^{\mathrm{eq}} = T/\Omega^+(k)$, and
$\widetilde{l}_{\mathbf{k}}^{\mathrm{eq}} = T/\Omega^-(k)$,
respectively. The condensates, therefore, strongly affect the
equilibrium distributions of the quasiparticles.

Using these equilibrium solutions, we can now derive a relation
between the occupation numbers $n_o$, $l_o$, and the total number of
particles $N_1$, $N_2$ and the energy $H$. For a finite sized
system, the Hamiltonian can be expressed in terms of the Fourier
series $\psi_1 = \frac{1}{\sqrt{V}} \sum_k a_k \exp(i\mathbf{k}
\cdot \mathbf{x})$, $\psi_2 = \frac{1}{\sqrt{V}} \sum_k b_k
\exp(i\mathbf{k} \cdot \mathbf{x})$. The Hamiltonian can then be
written as $H = H_o + H_2 + H_3 + H_4$ depending on how $a_o =
a_{k=0}$ and $b_o = b_{k=0}$, and non-zero modes enter the
expansion: $H_o = \frac{1}{2V} [ |a_o|^4 + |b_o|^4 +
2|a_o|^2(N_1-|a_o|^2) + 2|b_o|^2(N_2-|b_o|^2) ] + \frac{\alpha}{V}
\left[ |a_o|^2 N_2 + |b_o|^2 N_1 \right], \;\;\; H_2 = \sum_k^{'} [
(k^2 + \frac{|a_o|^2}{V}) a_{\mathbf{k}} a_{\mathbf{k}}^* +
\frac{1}{2V} (a_o^2 a_{\mathbf{k}}^* a_{\mathbf{-k}}^* + c.c.) ] +
\sum_k^{'} [ (k^2 + \frac{|b_o|^2}{V}) b_{\mathbf{k}}
b_{\mathbf{k}}^* + \frac{1}{2V} (b_o^2 b_{\mathbf{k}}^*
b_{\mathbf{-k}}^* + c.c.)] + \sum_k^{'} [ \frac{\alpha}{V}(a_o b_o
a_{\mathbf{k}}^* b_{\mathbf{-k}}^* + a_o b_o^* a_{\mathbf{k}}^*
b_{\mathbf{k}} + c.c.)], \;\;\; H_3 = \sum_{k_1,k_2,k_3}^{'}
[\frac{1}{2V} (2a_o a_{k_1} a_{k_2}^* a_{k_3}^* + 2b_o b_{k_1}
b_{k_2}^* b_{k_3}^* + c.c.) + \frac{\alpha}{V} (a_o b_{k_1}
a_{k_2}^* b_{k_3}^* + b_o a_{k_1} a_{k_2}^* b_{k_3}^* + c.c.)]
\delta_{\mathbf{k}_1-\mathbf{k}_2-\mathbf{k}_3}, \;\;\; H_4 =
\sum_{k_1,k_2,k_3,k_4}^{'} [\frac{1}{2V}(a_{k_1} a_{k_2} a_{k_3}^*
a_{k_4}^* + b_{k_1} b_{k_2} b_{k_3}^* b_{k_4}^*) + \frac{\alpha}{V}
(a_{k_1} b_{k_2} a_{k_3}^* b_{k_4}^*)]
\delta_{\mathbf{k}_1+\mathbf{k}_2-\mathbf{k}_3-\mathbf{k}_4}$.
$\sum_k^{'}$ denotes summation over $k$ but excluding the $k=0$
mode. To relate the equilibrium distributions obtained from the
kinetic equation (\ref{kn2}) to this form of the Hamiltonian, we
must diagonalise the quadratic term $H_2$. Rewriting the Hamiltonian
in terms of the basis $\mathcal{N}=(\xi_{\mathbf{k}},
\eta_{\mathbf{k}})^T$, the quadratic part takes the form $H_2 =
\sum_k' (\Omega^+(k) \xi_k \xi_k^* + \Omega^-(k) \eta_k \eta_k^*)$.
Now ensemble averaging the equations and using the equilibrium
distributions $\widetilde{n}_{\mathbf{k}}^{\mathrm{eq}}$, and
$\widetilde{l}_{\mathbf{k}}^{\mathrm{eq}}$ given above, we can
express the occupation numbers of the two gases in the new basis as
\begin{eqnarray}
N_1-n_o &=& T\sum_k' \frac{(u_{11}^{+^2}+u_{11}^{-^2})}{\Omega^+(k)} +
\frac{(u_{12}^{+^2}+u_{12}^{-^2})}{\Omega^-(k)}, \label{eqn_N1} \\
N_2-l_o &=& T\sum_k' \frac{(u_{21}^{+^2}+u_{21}^{-^2})}{\Omega^+(k)} +
\frac{(u_{22}^{+^2}+u_{22}^{-^2})}{\Omega^-(k)}. \label{eqn_N2}
\end{eqnarray}
The ensemble averaged Hamiltonian will have contributions from only
$H_0$, $H_2$, and $H_4$. Rewriting the resulting expression in the
new basis, we obtain: $\left< H \right> = E_0 + \sum_k' (\Omega_1(k)
\widetilde{n}_{\mathbf{k}}^{\mathrm{eq}} +\Omega_2(k)
\widetilde{l}_{\mathbf{k}}^{\mathrm{eq}}) = E_0 + 2T \sum_k' 1$.
$E_0=\frac{1}{2V}[N_1^2 + (N_1-n_o)^2 + N_2^2 + (N_2-l_o)^2] +
\frac{\alpha}{V}[N_1N_2]$ denotes the energy of the ground state.
Using either Eq.\ (\ref{eqn_N1})
or (\ref{eqn_N2}), we can eliminate the temperature in the
expression for $\left< H \right>$ to obtain
\begin{eqnarray}
\left< H \right> &=& E_0 + \frac{(N_1-n_o) \sum_k' 1}{\sum_k' \left(
\frac{(u_{11}^{+^2}+u_{11}^{-^2})}{\Omega^+(k)} +
\frac{(u_{12}^{+^2}+u_{12}^{-^2})}{\Omega^-(k)} \right)}, \nonumber \\
&=& E_0 + \frac{(N_2-l_o) \sum_k' 1}{\sum_k' \left( \frac{(u_{21}^{+^2}+u_{21}^{-^2})}
{\Omega^+(k)} + \frac{(u_{22}^{+^2}+u_{22}^{-^2})}{\Omega^-(k)} \right)}.
\end{eqnarray}
This equation provides two algebraic relations for the two unknowns
$n_o$ and $l_o$ given $H$, $N_1$, $N_2$. At intermediate energies,
we find that only one component condenses and the theory presented
above breaks down (in practice resulting in negative values of
$l_o$). In this part of the parameter space, we need to introduce
the ansatz $a_k = [\sqrt{n_o} \delta(k) + \widetilde{a}_k(t)]e^{-i
n_o t}$ only for the first component whilst assuming a purely
continuous spectrum for the second component. This results in a
coupled system of kinetic equations that are similar to Eqs.\
(\ref{eqn_kn1}) and (\ref{kn2}) above but where the first component
is governed by three-wave resonances whereas the second is governed
by four-wave resonances. These equations will have the equilibrium
distributions $\widetilde{n}_k = \frac{T}{\omega_{B}}$ and $l_k =
\frac{T}{k^2-\mu_2}$ where $\widetilde{n}_k$ is now defined as above
but with $l_o$ set to zero and $\omega_{B} = \sqrt{k^4 + 2n_o
k^2/V}$ is the classical single-component Bogoliubov dispersion
relation. Therefore, when only one component is condensed, we
obtain the expressions: $N_1-n_o = T\sum_k'
\frac{(k^2+n_o/V)}{\omega_{B}^2(k)}$ and $N_2 = \sum_k'
\frac{T}{k^2-\mu_2}$. The ensemble averaged Hamiltonian then takes
the form
\begin{eqnarray}
\left< H \right> &=& E_0 + \frac{(N_1-n_o) \sum_k' \left( 1 +
\frac{k^2}{k^2-\mu_2} \right)}{\sum_k' \frac{(k^2+n_o/V)}{\omega_{B}^2(k)} },
\nonumber \\
&=& E_0 + \frac{N_2 \sum_k' \left( 1 +
\frac{k^2}{k^2-\mu_2} \right)}{\sum_k' \frac{1}{k^2-\mu_2} }
\label{eqn_H_onecomp}
\end{eqnarray}
and now $E_0 = \frac{1}{2V}[N_1^2+(N_1-n_o)^2 + 2N_2^2 + 2\alpha N_1
N_2]$. This equation provides two algebraic expressions for the two unknowns
$n_o$ and $\mu_2$. We note that at the critical energy where the condensate in
the second component vanishes, we have $l_o = 0$ and $\mu_2 = 0$. At this
point, it can be shown that the equilibrium distributions given by
$\frac{T}{\Omega^+(k)}$ and $\frac{T}{\Omega^-(k)}$ reduce to
$\frac{T}{\omega_{B}(k)}$ and $\frac{T}{k^2}$ and the two solutions given
above match at the critical energy. This provides a solution for the
thermodynamic state that is uniformly valid over the entire range of energies.

\begin{figure}
\centering \epsfig{file=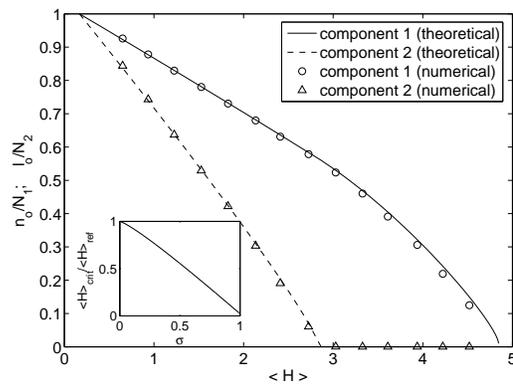,scale=0.375,angle=0.0}
\vspace{-0.53cm} \caption{\label{Fig_mf} Mass fractions as a
function of averaged total energy for $\frac{N_1}{V}=0.5$,
$\frac{N_2}{V}=0.25$, $\alpha = 0.1$; inset - non-dimensionalised
critical energy as a function of discrepancy parameter.}
\end{figure}

To verify the theory, we numerically solved the coupled system in a cubic
region with periodic boundary conditions using
a pseudo-spectral method with spatial resolution containing ($64^3$)
grid points and a time-step of 0.01.  The initial conditions were set
using the random phase approximation (see e.g.\ [4]).
To determine the properties of the system at equilibrium,
we assumed that the ergodic hypothesis applies and used time-averages from our
simulations to represent ensemble averages that arise in the theory presented
above.
Figure \ref{Fig_mf} presents results for the variation of the condensate mass
fractions with $\left< H \right>$. The results shown reveal remarkable
agreement with the theory presented over the entire range of energies. For very
large energies where only a small condensate mass fraction of the first component is
present, the theory deviates very slightly from the results of the
simulations. This occurs since many
particles are non-condensed in this region violating the assumption of a strong condensate
that is required for the theory.

Given the excellent agreement between the theory and predictions, we
can use Eq.\ (\ref{eqn_H_onecomp}) to determine how the critical
energy ($\left< H \right>_{\mathrm{crit}}$), at which condensation
in the second component ceases, varies with the discrepancy
parameter $\sigma = \frac{N_1-N_2}{N_1}$. The variation of $\left< H
\right>_{\mathrm{crit}}$, nondimensionalised with respect to the
value at $\sigma = 0$ ($\left< H \right>_{\mathrm{ref}}$), is shown
in the inset of Fig.\ \ref{Fig_mf}. The figure clearly illustrates
that $\left< H_{\mathrm{crit}} \right>$ deviates significantly from
its value at $\sigma = 0$ as the discrepancy parameter is increased
giving rise to a range of energies where only the first component
condenses.

In summary, we have derived a theoretical formulation of the thermodynamic
state governed by the coupled NLS equations for a classical system of two weakly
interacting waves in 3D.
Numerical simulations are in excellent quantitative agreement with the theory
that is based on the equilibrium solutions of the kinetic equations for the NLS
system of equations. The study presented here is relevant in quantifying 
condensation in a number of physical systems. Notable examples include 
mixtures of BEC gases at finite temperature and nonlinear optics
provided many modes are present that can be modeled semi-classically. 

The authors acknowledge support from EPSRC-UK under Grant No.\
EP/D032407/1 and Boris Svistunov for many useful discussions.

\end{document}